\documentclass{article}
\usepackage{amsmath,amssymb,graphicx,hyperref}
\usepackage{booktabs}
\usepackage{tabularx}
\usepackage[preprint]{spconf}
\usepackage{setspace}

\title{InconVAD: A Two-Stage Dual-Tower Framework for Multimodal Emotion Inconsistency Detection}

\name{Zongyi Li$^{1}$ \quad Junchuan Zhao$^{2}$ \quad Francis Bu Sung Lee$^{3}$ \quad Andrew Zi Han Yee$^{4}$}

\address{%
  \parbox{\linewidth}{\centering
    $^{1}$ Interdisciplinary Graduate Programme, Nanyang Technological University, Singapore\\
    $^{2}$ School of Computing, National University of Singapore, Singapore\\
    $^{3}$ College of Computing and Data Science, Nanyang Technological University, Singapore\\
    $^{4}$ Wee Kim Wee School of Communication and Information,
    Nanyang Technological University, Singapore
  }%
}

\copyrightnotice{%
  \begin{minipage}{\textwidth}
  \centering
  \footnotesize
    \textcopyright \ 20XX IEEE. Personal use of this material is permitted. Permission from IEEE must be obtained for all other uses, in any current or future media, including reprinting/republishing this material for advertising or promotional purposes, creating new collective works, for resale or redistribution to servers or lists, or reuse of any copyrighted component of this work in other works. 
  \end{minipage}
}
    
\begin{document}
\ninept

\maketitle

% \thanks{© 2025 IEEE. Personal use of this material is permitted. Permission from IEEE must be obtained for all other uses, in any current or future media, including reprinting/republishing this material for advertising or promotional purposes, creating new collective works, for resale or redistribution to servers or lists, or reuse of any copyrighted component of this work in other works.}

\begin{abstract}
Detecting emotional inconsistency across modalities is a key challenge in affective computing, as speech and text often convey conflicting cues. Existing approaches generally rely on incomplete emotion representations and employ unconditional fusion, which weakens performance when modalities are inconsistent. Moreover, little prior work explicitly addresses inconsistency detection itself. We propose InconVAD, a two-stage framework grounded in the Valence–Arousal–Dominance (VAD) space. In the first stage, independent uncertainty-aware models yield robust unimodal predictions. In the second stage, a classifier identifies cross-modal inconsistency and selectively integrates consistent signals. Extensive experiments show that InconVAD surpasses existing methods in both multimodal emotion inconsistency detection and modeling, offering a more reliable and interpretable solution for emotion analysis.
\end{abstract}
\begin{keywords}
Multimodal emotion inconsistency detection, Affective computing, Cross-modal representation learning, Multimodal emotion analysis
\end{keywords}
% \vspace{1em}
% \noindent \textbf{© 2025 IEEE.} Personal use of this material is permitted. Permission from IEEE must be obtained for all other uses, in any current or future media, including reprinting/republishing this material for advertising or promotional purposes, creating new collective works, for resale or redistribution to servers or lists, or reuse of any copyrighted component of this work in other works.
\section{Introduction}
\label{sec:intro}
\begin{figure*}[t!]
    \centering
    \includegraphics[width=0.85\linewidth]{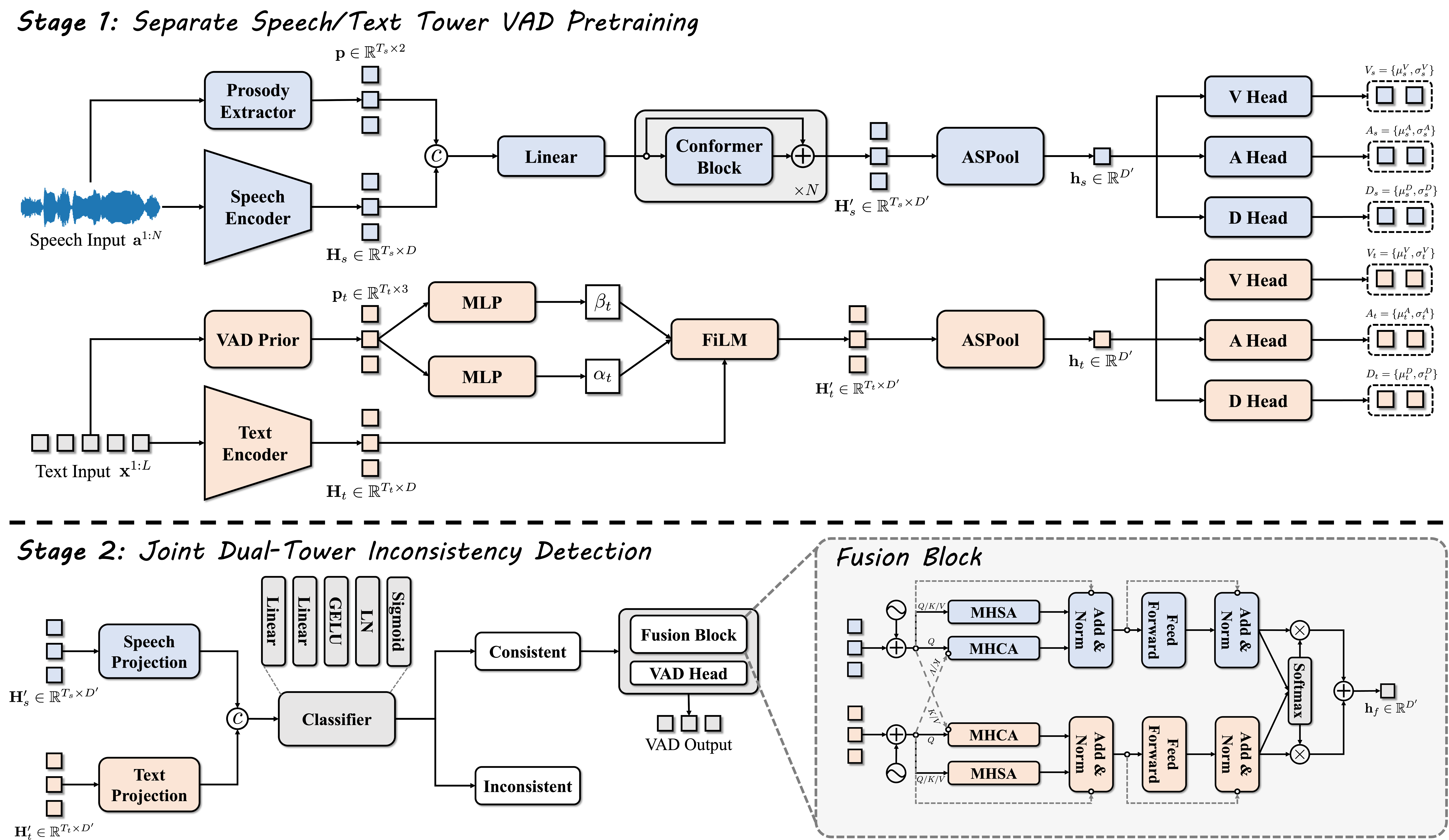}
    \caption{Overview of the proposed InconVAD framework with two phases: Phase A builds speech and text towers for unimodal VAD estimation, while Phase B employs an inconsistency detection head and a fusion module for selective integration.}
    \label{fig:overall}
\end{figure*} 
With the rapid development of human–computer interaction systems, accurately modeling emotions across multiple modalities has become a central challenge in affective computing \cite{Xu2024CrossTask}. A key difficulty lies in the fact that speech and text do not always convey consistent affective states. Such inconsistencies may reflect complex psychological mechanisms, social strategies, or even clinical conditions \cite{Su2024Investigating}. Detecting and quantifying multimodal emotion inconsistency therefore requires a reliable framework that directly compares emotional representations across modalities, rather than relying solely on unimodal cues or generic fusion strategies.

Current approaches to cross-modal emotion analysis face two fundamental challenges. First, they often rely on inadequate emotional representation models, such as discrete emotion categories, which fail to capture the nuance and continuity of real-world affective expressions. Existing studies on emotion inconsistency at the categorical level typically reduce the task to comparing emotion labels across modalities. For example, \cite{Wang2023Multimodal} introduced the Multimodal Cross-Attention Bayesian Network (MCABN), which employed attention mechanisms to identify inconsistencies between images and text on social media, using category-based labels with pseudo-label guidance to resolve conflicts. Similarly, \cite{Su2024Investigating} proposed a framework for detecting Acoustic–Text Emotion Inconsistency (ATEI) in depression diagnosis by categorizing emotions into three classes (positive, negative, neutral). While such label-based methods are computationally efficient and offer clear interpretability, they overlook variations in emotional intensity. As a result, they provide only a coarse quantification of inconsistency, leading to the loss of fine-grained affective information in practical applications that require detailed emotional analysis.

Second, most multimodal emotion recognition methods are built upon the implicit assumption that modalities such as speech and text are emotionally congruent \cite{ tsai2019multimodal}. When this assumption is violated, their fusion strategies—typically averaging or concatenation—produce vague intermediate representations that dilute the literal sentiment expressed in text and obscure the authentic emotional cues in speech \cite{pramanick2022multimodal, wang2023cross}. Moreover, these models generally lack mechanisms for uncertainty awareness, treating all inputs as equally reliable and failing to assign greater weight to the clearer or more trustworthy modality when discrepancies arise. The absence of explicit modeling for emotion inconsistency thus remains a critical deficiency in current multimodal emotion recognition research, underscoring the need for frameworks that directly address inconsistency rather than treating it as a byproduct of fusion \cite{pan2020modeling}.
% \cite{Fantini2023Consistency}

To address these issues, we propose InconVAD, a two-stage framework grounded in the Valence–Arousal–Dominance (VAD) space. In the first stage, modality-specific towers independently predict VAD values with uncertainty-aware estimation, providing robust and comparable unimodal representations. In the second stage, a classifier explicitly detects cross-modal inconsistency, while a gated fusion module selectively integrates predictions only for consistent pairs. This design prevents representation collapse in cases such as sarcasm and preserves modality-specific cues that would otherwise be lost. Extensive experiments demonstrate that InconVAD surpasses existing methods in both multimodal emotion inconsistency detection and modeling, while offering greater interpretability through modality-specific VAD predictions.

% \cite{Mehrabian1996PleasureAA}
\section{METHODOLOGY}
\label{sec:method}

As illustrated in Fig.~\ref{fig:overall}, the proposed \textbf{InconVAD} framework operates in two stages. Stage 1 employs two unimodal towers for VAD pretraining: the speech tower and the text tower process raw speech and text inputs, respectively, and output modality-specific representations 
$\mathbf{h}_s$ and $\mathbf{h}_t$, together with VAD means $\boldsymbol{\mu}_s, \boldsymbol{\mu}_t$ and log-variances $\log \boldsymbol{\sigma}^2_s, \log \boldsymbol{\sigma}^2_t$. 
Stage 2 comprises an inconsistency detection head and a gated fusion tower. It takes the outputs of Stage~1, where the detection head predicts an inconsistency score $p_{\text{inc}}$, and the fusion tower, activated only for consistent pairs, generates a fused representation $\mathbf{h}_f$ and the final VAD prediction $\mathbf{y}_f$. 
The following subsections describe each component in detail.

\subsection{Phase A: Unimodal VAD Pretraining}\label{subsec:phase_a}
\subsubsection{Speech Tower}\label{subsubsec:speech_tower}
The speech tower extracts both acoustic and prosodic cues for reliable VAD estimation. 
We employ a pre-trained Wav2Vec2-base model $f_{SE}(\cdot)$ \cite{baevski2020wav2vec} to generate frame-level acoustic embeddings $\mathbf{H}_s \in \mathbb{R}^{T_s \times D}$, and a prosody extractor $f_{PE}(\cdot)$ to compute pitch and energy features $\mathbf{p} \in \mathbb{R}^{T_s \times 2}$ \cite{zhao25d_interspeech}. 
These complementary features are concatenated and projected through a linear layer to form the input $\mathbf{H}_{\mathrm{in}} \in \mathbb{R}^{T_s \times D'}$, ensuring that prosodic variation is explicitly incorporated into the acoustic space. 
The sequence is then processed by two Conformer blocks \cite{gulati2020conformer}, which integrate multi-head self-attention, convolutional modules, and Macaron-style feed-forward layers to capture both local dynamics and long-range dependencies. 
The resulting contextualized features $\mathbf{H}'_s$ are aggregated by an ASPool module \cite{okabe2018attentive}, producing a fixed-dimensional utterance-level embedding $\mathbf{h}_s \in \mathbb{R}^{D'}$. 
Finally, prediction heads output per-dimension means $\boldsymbol{\mu}_s \in \mathbb{R}^3$ and log-variances $\log\boldsymbol{\sigma}^2_s \in \mathbb{R}^3$, yielding uncertainty-aware unimodal estimates:  
\begin{equation}
\begin{aligned}
\mathbf{H}_{\mathrm{in}} &= \mathrm{Linear}([f_{SE}(\mathbf{a}), f_{PE}(\mathbf{a})]), \\
\mathbf{h}_s &= \mathrm{ASPool}(\mathrm{Conformer}(\mathbf{H}_{\mathrm{in}})), \\
(\boldsymbol{\mu}_s, \log\boldsymbol{\sigma}^2_s) &= f_{h}(\mathbf{h}_s).
\end{aligned}
\end{equation}

\subsubsection{Text Tower}\label{subsubsec:text_tower}
The text tower captures semantic and lexical-level affective cues, complementing the speech tower. 
A RoBERTa-base encoder $f_{TE}(\cdot)$ \cite{liu2019roberta} maps tokenized inputs $\mathbf{x} \in \mathcal{V}_{\rm text}^L$ to contextual embeddings $\mathbf{H}_t \in \mathbb{R}^{T_t \times D}$, which encode rich semantic and syntactic information. 
To explicitly inject affective knowledge, we employ the NRC VAD Lexicon v2 $f_{\mathrm {Prior}}(\cdot)$ \cite{mohammad2025nrc} to derive token-level prior vectors $\mathbf{p}_t \in \mathbb{R}^{T_t \times 3}$, representing valence, arousal, and dominance values. 
These priors are integrated with the encoder outputs using a FiLM layer \cite{perez2018film}, producing gated representations $\mathbf{H}_t'$ that remain dimensionally consistent while incorporating explicit affective supervision. 
An ASPool module then aggregates $\mathbf{H}_t'$ into an utterance-level embedding $\mathbf{h}_t \in \mathbb{R}^{D'}$, which is passed to prediction heads to produce $\boldsymbol{\mu}_t \in \mathbb{R}^3$ and $\log\boldsymbol{\sigma}^2_t \in \mathbb{R}^3$. 
This process yields modality-specific textual predictions that align with the speech tower outputs in the shared VAD space:  
\begin{equation}
\begin{aligned}
\mathbf{H}_t' &= \mathrm{FiLM}(f_{TE}(\mathbf{x}), f_{\mathrm{Prior}}(\mathbf{x})), \\
\mathbf{h}_t &= \mathrm{ASPool}(\mathbf{H}_t'), \\
(\boldsymbol{\mu}_t, \log\boldsymbol{\sigma}^2_t) &= f_{h}(\mathbf{h}_t).
\end{aligned}
\end{equation}

\subsubsection{Training Strategy}\label{subsubsec:train_a}

To optimize the unimodal towers, we adopt a heteroscedastic regression framework, where the prediction uncertainty is explicitly modeled through variance estimation. The training objective is the Gaussian Negative Log-Likelihood (NLL) of the ground-truth labels \cite{NIPS2017_2650d608}. For a given modality $m \in \{s, t\}$ (speech or text) and dimension $k \in \{V, A, D\}$, the loss is defined as:
\begin{equation}
    \mathcal{L}_{\text{NLL}}^{(m,k)} = \frac{(y_m^{k} - \mu_m^{k})^2}{2\,{\sigma_m^{k}}^2} + \tfrac{1}{2}\log({\sigma_m^{k}}^2),
\end{equation}
where $y_m^{k}$ denotes the ground-truth label, and $\mu_m^{k}$ and ${\sigma_m^{k}}^2$ are the predicted mean and variance, respectively. 

\subsection{Phase B: Inconsistency Detection with Gated Fusion}\label{subsec:phase_b}

\subsubsection{Inconsistency Detection Classifier}\label{subsubsec:classifier}
The inconsistency classifier is designed to assess cross-modal inconsistency without modifying the unimodal feature extractors trained in Phase~A. It operates on speech and text representations, $\mathbf{H}_s'$ and $\mathbf{H}_t'$, and determines whether they convey consistent affective states.  

To align the modalities, the representations are first projected into a shared latent space through lightweight projectors $f_{SP}(\cdot)$ and $f_{TP}(\cdot)$, producing $\tilde{\mathbf{H}}_s, \tilde{\mathbf{H}}_t \in \mathbb{R}^{T \times D'}$. These projections normalize the features and reduce domain gaps between modalities. The two projected sequences are then concatenated to form a joint representation $\tilde{\mathbf{H}} \in \mathbb{R}^{T \times 2D'}$, which is passed to a binary classifier $f_C(\cdot)$. The classifier consists of two linear layers with GELU activation, followed by LayerNorm and a Sigmoid output, yielding the predicted inconsistency score $p_{\text{inc}} \in [0,1]$.  
The overall classification process can be expressed as: 
\begin{equation}
    p_{\text{inc}} = f_C\!\left([f_{SP}(\mathbf{H}_s'), f_{TP}(\mathbf{H}_t')]\right).
\end{equation}

\subsubsection{Fusion Module}\label{subsubsec:fusion_tower}
Our fusion module is an end-to-end architecture that integrates speech and text features to predict VAD emotional dimensions. 
Building on the outputs of the speech and text towers, the projected sequences are passed into a  cross-modal fusion module. 
Inspired by \cite{gu2022mm} and \cite{ma2024transformer}, we design a Transformer block that jointly models intra- and inter-modal dependencies. 
Specifically, multi-head self-attention (MHSA) is applied to capture intra-modal relationships ($\text{speech}\!\to\!\text{speech}$ and $\text{text}\!\to\!\text{text}$), while multi-head cross-attention (MHCA) enables cross-modal interactions ($\text{speech}\!\to\!\text{text}$ and $\text{text}\!\to\!\text{speech}$). 
The outputs are subsequently processed with LayerNorm (LN) and feed-forward networks (FFN) to obtain modality-specific contextual representations $\mathbf{f}_s, \mathbf{f}_t \in \mathbb{R}^{L \times D'}$.  

To dynamically integrate information across modalities, we utilize the gated multimodal fusion mechanism. 
Each modality is first projected through a learnable weight matrix and then normalized via a softmax function applied along the modality axis. 
This produces element-wise gates $\mathbf{g}_s, \mathbf{g}_t \in \mathbb{R}^{T \times 1}$, which adaptively control the contribution of each modality at every time step. 
The final fused representation $\mathbf{h}_f \in \mathbb{R}^{T \times D'}$ is obtained as the weighted combination of modality-specific features. 
The entire process can be formulated as:
\begin{equation}
\begin{aligned}
\mathbf{f}_s' &= \mathrm{LN}\!\big(\tilde{\mathbf{H}}_s 
    + \mathrm{MHSA}(\tilde{\mathbf{H}}_s) 
    + \mathrm{MHCA}(\tilde{\mathbf{H}}_s,\tilde{\mathbf{H}}_t)\big), \\
\mathbf{f}_t' &= \mathrm{LN}\!\big(\tilde{\mathbf{H}}_t 
    + \mathrm{MHSA}(\tilde{\mathbf{H}}_t) 
    + \mathrm{MHCA}(\tilde{\mathbf{H}}_t,\tilde{\mathbf{H}}_s)\big), \\
\mathbf{f}_s &= \mathrm{LN}(\mathrm{FFN}(\mathbf{f}_s') + \mathbf{f}_s'), \quad
\mathbf{f}_t = \mathrm{LN}(\mathrm{FFN}(\mathbf{f}_t') + \mathbf{f}_t'), \\
[\mathbf{g}_s, \mathbf{g}_t] &= \mathrm{softmax}\big([\mathbf{f}_s \cdot \mathbf{W}_s,\; \mathbf{f}_t \cdot \mathbf{W}_t], \\
\mathbf{h}_f &= \mathbf{g}_s \odot \mathbf{f}_s + \mathbf{g}_t \odot \mathbf{f}_t ,
\end{aligned}
\end{equation}
where $\mathbf{W}_s, \mathbf{W}_t \in \mathbb{R}^{D' \times D'}$ are learnable projections, 
$[\mathbf{g}_s, \mathbf{g}_t]$ are learned gates for speech and text, and $\odot$ denotes the element-wise product with broadcasting along the feature dimension.

\subsubsection{Training Strategy}\label{subsubsec:train_b}
The inconsistency classifier is trained with a joint objective that combines classification accuracy and geometric regularization of the latent space. 
The primary term is the Binary Cross-Entropy loss, $\mathcal{L}_{\text{BCE}}$, which compares the predicted inconsistency score $p_{\text{inc}}$ against the ground-truth label \cite{hopfield1987neural}. 
To further structure the latent space, we introduce a margin-based auxiliary loss, $\mathcal{L}_{\text{margin}}$, applied to the projected representations $\tilde{\mathbf{H}}_s$ and $\tilde{\mathbf{H}}_t$:  
\begin{equation}
    \mathcal{L}_{\text{margin}} = y \cdot d^2 + (1-y) \cdot \max(0, m - d)^2,
\end{equation}
where $d = \lVert \tilde{\mathbf{H}}_s - \tilde{\mathbf{H}}_t \rVert_2$ denotes the Euclidean distance. 
This term pulls consistent pairs ($y=1$) closer in the latent space, while enforcing a minimum margin $m$ between inconsistent pairs ($y=0$) \cite{hadsell2006dimensionality}. 
The final training objective combines the two losses with a weighting factor $\lambda_{\text{margin}}$:  
\begin{equation}
    \mathcal{L}_{\text{CLS}} = \mathcal{L}_{\text{BCE}} + \lambda_{\text{margin}} \cdot \mathcal{L}_{\text{margin}}.
\end{equation}

For the fusion module, training is guided by two complementary objectives. 
On labeled data with available VAD annotations, we apply the same Gaussian NLL loss as in Stage~1. 
For unlabeled pairs without VAD ground truth, we introduce a selective agreement loss, which encourages the fused prediction $(\boldsymbol{\mu}_f,\log\boldsymbol{\sigma}_f^2)$ to agree with a Gaussian target derived from the unimodal outputs, defined as  
\begin{equation}
\mathcal{L}_{\text{agree}}
= \sum_{k\in\{V,A,D\}}
\left[
\frac{(\mu_f^k-\mu_{\text{agree}}^k)^2}{2{\sigma_{\text{agree}}^{k}}^2}
+ \tfrac{1}{2}\log{\sigma_{\text{agree}}^{k}}^2
\right],
\end{equation}
where $\mu_{\text{agree}}^k=\tfrac{\mu_s^k/{\sigma_s^{k}}^2+\mu_t^k/{\sigma_t^{k}}^2}{1/{\sigma_s^{k}}^2+1/{\sigma_t^{k}}^2}$ and ${\sigma_{\text{agree}}^{k}}^2=\tfrac{1}{1/{\sigma_s^{k}}^2+1/{\sigma_t^{k}}^2}$. 
The overall training objective of the fusion tower combines the supervised and selective terms:  
\begin{equation}
\mathcal{L}_{\text{Fusion}} = \mathcal{L}_{\text{NLL}} + \lambda_{\text{agree}} \cdot \mathcal{L}_{\text{agree}}.
\end{equation}

\section{EXPERIMENTAL SETUPS}\label{sec:exp_setups}
\subsection{Datasets}\label{subec:dataset}
For Phase~A, we use the IEMOCAP \cite{busso2008iemocap} dataset for the speech tower and the EmoBank \cite{buechel2017emobank} dataset for the text tower, both with dimensional Valence–Arousal–Dominance (VAD) annotations. To ensure label comparability across datasets, we apply a parametric Beta Cumulative Distribution Function (CDF) transform that maps each original label $v$ into an aligned value $v'$ in a shared target distribution. A source value $v$ is first normalized to $[0,1]$, then mapped to its quantile $u = F_{src}(v)$ using the source CDF, and finally aligned by applying the target inverse CDF, $v' = F_{tgt}^{-1}(u)$. The aligned labels $v'$ are used as training targets, while during evaluation the model predictions $\hat{v}'$ are mapped back through the inverse transform to obtain $\hat{v}$ for comparison against the original labels $v$. The Beta-CDF process can be formulated as:
\begin{equation}
v' = F_{tgt}^{-1}(F_{src}(v)).
\end{equation}

For Phase B, to train the inconsistency classifier, we use IEMOCAP \cite{busso2008iemocap} dataset and the EmoV-DB \cite{adigwe2018emotional} dataset to construct binary-labeled data pairs. Consistency pairs (label $y=1$) include speech-text pairs from IEMOCAP and neutral emotion speech-text pairs from EmoV-DB. Inconsistent pairs ($y=0$) are generated from EmoV-DB by pairing neutral transcripts with non-neutral speech recordings of the same utterance ID. In addition, to train the fusion tower, we use only consistency pairs ($y=1$), as cross-modal fusion is meaningful only when the two modalities are emotionally aligned. The tower is trained to integrate the unimodal predictions into a single fused VAD output $\mathbf{y}_f$, providing a unified estimate rather than separate outputs for each modality.

% \subsubsection{Cross-Dataset VAD Label Alignment}\label{subsubsec:data_align}

\begin{table}[t]
\centering
\footnotesize
\caption{Comparison of unimodal and fusion towers using Concordance Correlation Coefficient (CCC).}
\vspace{0.1cm}
\label{tab:ccc-iemocap}
\setlength{\tabcolsep}{5pt}
\renewcommand{\arraystretch}{1} 

\begin{tabular}{l *{4}{>{\centering\arraybackslash}m{0.8cm}}}
\toprule
\textbf{Method} & \textbf{V} & \textbf{A} & \textbf{D} & \textbf{Avg} \\
\midrule
\textbf{Ours (Speech Tower)} & \textbf{0.639} & \textbf{0.669} & \textbf{0.541} & \textbf{0.616} \\
\textbf{Ours (Text Tower)}   & \textbf{0.784} & \textbf{0.419} & \textbf{0.443} & \textbf{0.549} \\
\midrule
Dimensional MTL~\cite{atmaja2020dimensional} & 0.446 & 0.594 & 0.485 & 0.508 \\
Two-stage SVM~\cite{atmaja2021two}         & 0.595 & 0.601 & 0.499 & 0.565 \\
RL-MT~\cite{srinivasan2022representation} & 0.648 & 0.668 & 0.537 & 0.618 \\
MFCNN14~\cite{triantafyllopoulos2023multistage} & 0.714 & 0.639 & 0.575 & 0.642 \\
W2v2-b + BERT-b + L~\cite{zhang2024mersa} & 0.625 & 0.661 & 0.570 & 0.618 \\
\midrule
\textbf{Ours (Fusion Tower)} & \textbf{0.741} & \textbf{0.644} & \textbf{0.586} & \textbf{0.657} \\
\bottomrule
\end{tabular}
\end{table}

\subsection{Implementation Details}\label{subsec:implement_details}

In Phase~A, both towers use pretrained backbones, Wav2Vec2-base for speech and RoBERTa-base for text (hidden size 768), followed by projection layers of dimension 256. Training is performed with a batch size of 16 for up to 50 epochs with early stopping (patience = 5). Optimization uses AdamW with learning rates of $2\times10^{-5}$ for the backbone and $1\times10^{-4}$ for the heads, combined with a cosine schedule and 10\% warm-up. Weight decay is set to 0.01. The minimum variance of $2\times10^{-3}$ is used by the heteroscedastic Gaussian NLL loss function. We use the concordance correlation coefficient (CCC) as the evaluation metric. To avoid data leakage, we use a speaker-independent split for IEMOCAP. All utterances from each of the ten speakers go to a single partition with an 8/1/1 train/validation/test ratio via group-based splitting. For EmoBank, we follow the official train/validation/test split annotated in the corpus.

For Phase~B inconsistency detection, both the speech and text towers are kept frozen, and optimization is performed solely on the classifier head. Each pair of data forms the input after modality-specific linear projections to a 256-dimensional space. We use a batch size of 32 and train for up to 50 epochs with early stopping (patience = 5). Optimization uses AdamW on classifier parameters, with a learning rate of $1\times10^{-3}$ and weight decay $0.01$. The loss combines binary cross-entropy and a margin term (margin $m=0.9$, $\lambda=0.15$). For fusion tower, we keep the batch size at 16 and train for up to 50 epochs. Optimization uses AdamW with learning rate  $1\times10^{-4}$, weight decay 0.01, and a cosine schedule with 10\% warm-up. We use CCC as the evaluation metric. For data split, we use the same speaker-independent split as in Phase~A for IEMOCAP dataset, while EmoV-DB utterances are partitioned into 8/1/1 train/validation/test sets.
\section{EXPERIMENTAL RESULTS}
% \subsection{Comparison with SOTA Approaches}

In Phase~A, our unimodal speech and text towers obtain average CCCs of 0.616 on IMEOCAP dataset and 0.549 on Emobank dataset, respectively. The fusion tower achieves 0.657, surpassing the existing state-of-the-art fusion models, including Dimensional MTL~\cite{atmaja2020dimensional}, Two-stage SVM ~\cite{atmaja2021two}, RL-MT ~\cite{srinivasan2022representation}, MFCNN14 ~\cite{triantafyllopoulos2023multistage}, and W2v2-b + BERT-b + L ~\cite{zhang2024mersa}, as shown in Table~\ref{tab:ccc-iemocap}. The consistent gains across Valence, Arousal, and Dominance highlight the complementary strengths of speech and text tower, and the effectiveness of our transformer blocks and gated fusion mechanism.

For the inconsistency detection task, our classifier achieves the best performance across reported metrics. As shown in Table~\ref{tab:baseline-compare}, On the test set, it attains an accuracy of 92.3\% and an F1-score of 92.2\%, surpassing prior methods (SVM \cite{castro2019mustard}, ATEI \cite{Su2024Investigating}). Precision and recall are likewise competitive, confirming that the leakage-free training protocol and composite loss design enable clear separation between consistent and inconsistent pairs. The decision threshold $\tau^*$ was fixed based on validation by maximizing Youden’s $J$, ensuring fair evaluation.

\begin{table}[t]
\centering
\footnotesize
\caption{Comparison with SOTA methods on emotional inconsistency detection.}
\vspace{0.1cm}
\label{tab:baseline-compare}
\setlength{\tabcolsep}{5pt}
\renewcommand{\arraystretch}{1}

\begin{tabular}{l *{4}{>{\centering\arraybackslash}m{1.2cm}}}
\toprule
\textbf{Method} & \textbf{Accuracy} & \textbf{F1-Score} & \textbf{Precision} & \textbf{Recall} \\
\midrule
SVM~\cite{castro2019mustard} & 85.7 & 86.4 & 80.3 & 93.6 \\
ATEI~\cite{Su2024Investigating} & 83.4 & 83.6 & 82.2 & 85.0 \\
\textbf{Ours (Classifier)}       & \textbf{92.3} & \textbf{92.2} & \textbf{93.6} & \textbf{90.9} \\
\bottomrule
\end{tabular}
\end{table}

\begin{table}[t]
\centering
\footnotesize
\caption{Ablation results for the speech, text, and fusion towers, with all metrics reported in CCC.}
\vspace{0.1cm}
\label{tab:ccc-abla}
\setlength{\tabcolsep}{5pt}
\renewcommand{\arraystretch}{1}

\begin{tabular}{l *{4}{>{\centering\arraybackslash}m{0.8cm}}}
\toprule
\textbf{Method} & \textbf{V} & \textbf{A} & \textbf{D} & \textbf{Avg} \\
\midrule
w/o Prosody Injection       & 0.608 & 0.634 & 0.514 & 0.585 \\
w/o Conformer Blocks        & 0.592 & 0.661 & 0.499 & 0.584 \\
w/o Attentive Statistics Pooling       & 0.627 & 0.654 & 0.556 & 0.612 \\
\midrule
\textbf{Ours (Speech Tower)} & \textbf{0.639} & \textbf{0.669} & \textbf{0.541} & \textbf{0.616} \\
\midrule
w/o Affect Prior Gating     & 0.776 & 0.447 & 0.406 & 0.543 \\
w/o Attentive Statistics Pooling       & 0.778 & 0.426 & 0.435 & 0.546 \\
\midrule
\textbf{Ours (Text Tower)} & \textbf{0.784} & \textbf{0.419} & \textbf{0.443} & \textbf{0.549} \\
\midrule
w/o Transformer Block       & 0.706 & 0.664 & 0.554 & 0.641 \\
w/o Gated Multimodal Fusion & 0.720 & 0.622 & 0.534 & 0.625 \\
\midrule
\textbf{Ours (Fusion Tower)} & \textbf{0.741} & \textbf{0.644} & \textbf{0.586} & \textbf{0.657} \\
\bottomrule
\end{tabular}
\end{table}

% \begin{table}[t]
% \centering
% \footnotesize
% \caption{Comparison with SOTA methods on emotional inconsistency detection.}
% \vspace{0.1cm}
% \label{tab:baseline-compare}
% \setlength{\tabcolsep}{5pt}
% \renewcommand{\arraystretch}{1}

% \begin{tabular}{l *{4}{>{\centering\arraybackslash}m{1.2cm}}}
% \toprule
% \textbf{Method} & \textbf{Accuracy} & \textbf{F1-Score} & \textbf{Precision} & \textbf{Recall} \\
% \midrule
% SVM~\cite{castro2019mustard} & 85.7 & 86.4 & 80.3 & 93.6 \\
% ATEI~\cite{Su2024Investigating} & 83.4 & 83.6 & 82.2 & 85.0 \\
% \textbf{Ours (Classifier)}       & \textbf{91.2} & \textbf{90.1} & \textbf{89.6} & \textbf{90.5} \\
% \bottomrule
% \end{tabular}
% \end{table}

We conduct a series of ablation studies to validate the necessity our architectural design as shown in Table~\ref{tab:ccc-abla}. For the speech tower, removing the Conformer blocks or prosody injection substantially reduces average CCC, highlighting the importance of temporal modeling and prosodic cues. Eliminating ASPool module also leads to a consistent drop, confirming its role in emphasizing salient acoustic features. For text tower, removing the affect prior gating decreases average CCC from 0.549 to 0.543, validating the benefit of injecting lexical affective knowledge. Similarly, discarding ASPool module lowers overall performance. For the Fusion Tower, removing Transformer block or the gated multimodal fusion mechanism degrades the average CCC from 0.657 to 0.641 and 0.625, respectively. These results confirm that modeling mutual interactions and dimension-specific modality weighting are both critical for robust cross-modal integration.

% \begin{table}[t]
% \centering
% \footnotesize % <-- Use this command for a smaller size
% \caption{Ablation study for the Speech, Text, and Fusion Towers. All metrics are CCC.}
% \vspace{0.1cm}
% \label{tab:ccc-abla}
% \setlength{\tabcolsep}{5pt}
% \renewcommand{\arraystretch}{1.1}
% \begin{tabular}{l c c c c}
% \toprule
% \textbf{Method} & \textbf{V} & \textbf{A} & \textbf{D} & \textbf{Avg.} \\
% \midrule
% w/o Prosody Injection & 0.608 & 0.634 & 0.514 & 0.586 \\
% w/o Conformer & 0.592 & 0.661 & 0.499 & 0.584 \\
% w/o Attentive Pooling & 0.627 & 0.654 & 0.556 & 0.612 \\
% \midrule
% \textbf{Baseline (Speech Tower)} & \textbf{0.639} & \textbf{0.669} & \textbf{0.541} & \textbf{0.616} \\
% \bottomrule
% w/o Affect Prior Gating & 0.776 & 0.447 & 0.406 & 0.544 \\
% w/o Attentive Pooling & 0.778 & 0.426 & 0.435 & 0.546 \\
% \midrule
% \textbf{Baseline (Text Tower)} & \textbf{0.784} & \textbf{0.419} & \textbf{0.443} & \textbf{0.549} \\
% \bottomrule
% w/o Transformer Block  & 0.706 & 0.664 & 0.554 & 0.641 \\
% w/o Gated Multimodal Fusion & 0.720 & 0.622 & 0.534 & 0.625 \\
% \midrule
% \textbf{Baseline (Fusion Tower)} & \textbf{0.741} & \textbf{0.644} & \textbf{0.586} & \textbf{0.657} \\
% \bottomrule
% \end{tabular}
% \end{table}

\section{CONCLUSIONS}

In this study, we propose InconVAD, a cross-modal emotion inconsistency detection framework grounded in a shared three-dimensional VAD emotion space. The framework produces interpretable and comparable VAD predictions from both speech and text, enabling effective inconsistency detection across modalities. This work establishes a foundation for building more emotionally intelligent and trustworthy human–computer interaction systems in real-world applications. Furthermore, our study highlights the importance of explicitly modeling cross-modal inconsistencies rather than assuming unimodal agreement, paving the way for more reliable multimodal affective computing systems.

\vfill\pagebreak

% \section{REFERENCES}
% \label{sec:refs}

% List and number all bibliographical references at the end of the
% paper. The references can be numbered in alphabetic order or in
% order of appearance in the document. When referring to them in
% the text, type the corresponding reference number in square
% brackets as shown at the end of this sentence \cite{C2}. An
% additional final page (the fifth page, in most cases) is
% allowed, but must contain only references to the prior
% literature.

% Please follow the IEEE Citation Guidelines, \url{https://ieee-dataport.org/sites/default/files/analysis/27/IEEE\%20Citation\%20Guidelines.pdf} for formatting of references.

% \bibliographystyle{IEEEbib}
\bibliographystyle{IEEEbib}
% \bibliography{strings,refs}
{\ninept        
% \setstretch{0.9}  
\bibliography{InConVAD}
}
\end{document}